\documentclass[twocolumn,twoside,slac_two]{revtex4}
\usepackage{graphicx}
\usepackage{fancyhdr}
\newcommand{\pt}{\rm p_{\rm T}}
\newcommand{\ptjet}{\rm p_{\rm T}^{\rm jet}}
\newcommand{\etjet}{\rm E_{\rm T}^{\rm jet}}

\newcommand{\kt}{\rm k_{\rm T}}
\newcommand{\yjet}{\rm y^{\rm jet}}
\newcommand{\ptg}{\rm p_{\rm T}^{\gamma}}
\begin{document}

\title{High $\pt$ Jet Physics}

\author{M. Mart\'\i nez-P\'erez}
\affiliation{IFAE, Barcelona, Spain}

\begin{abstract}
In this contribution, a comprehensive review of the main aspects of high $\pt$ jet physics
in Run II at the Tevatron is presented. Recent measurements on inclusive jet production are 
discussed using different jet algorithms and covering a wide region of jet transverse momentum
and jet rapidity. Several measurements, sensitive to a proper description of soft gluon radiation
and the underlying event in hadron collisions, are shown.  Finally,  high $\pt$  prompt photon 
measurements and studies on the production of  electroweak bosons in association with jets 
in the final state are discussed. 

\end{abstract}

\maketitle

\thispagestyle{fancy}

\section{INCLUSIVE JET PRODUCTION}

The measurement of the inclusive jet cross section in $p\overline{p}$ collisions at 
$\sqrt{s} = 1.96 \ \rm TeV$  constitutes a stringent test 
of perturbative QCD (pQCD) predictions over almost nine orders of magnitude.
The increased center-of-mass energy and integrated luminosity in Run II at the Tevatron, 
compared to Run I, allows to extend the measured jet cross section to jets with transverse 
momentum, $\ptjet$,  above $650 \ {\rm GeV/c}$, and to search for
signals of quark compositeness down to $\sim 10^{-19} \rm m$.

The pQCD calculations are written as matrix elements, describing the hard interaction
between partons, convoluted
with parton density functions (PDFs) in the proton and antiproton that 
require input from the experiments. 
Inclusive jet cross section  measurements from Run I at the Tevatron~\cite{d0runI,runIjet}, performed in 
different jet rapidity regions, 
have been used  to partially constrain the gluon distribution in the proton.
The pQCD  predictions are affected by the still limited knowledge on the gluon 
PDF,  which translates into a big uncertainty on the theoretical cross sections at high $\ptjet$.

The hadronic final states in hadron-hadron collisions are characterized by the presence of
soft contributions (the so-called {\it{underlying event}}) from initial-state gluon radiation 
and multiple parton interactions between remnants, in addition to the jets of hadrons originated 
by the hard interaction. A proper comparison with pQCD predictions at the parton level 
requires an adequate modeling of these soft contributions which become important at low $\ptjet$.

In Run II, both CDF and D0~experiments explore new 
jet algorithms following the theoretical work that indicates that the 
cone-based jet algorithm employed in Run I 
is not infrared safe and compromises a future 
meaningful comparison with pQCD calculations at NNLO.

 The CDF experiment recently published
results~\cite{ktprl,runIIjet} on inclusive jet production using the $\kt$~\cite{ktalgo,soper}  
and midpoint~\cite{midpoint} algorithms for jets with
$\ptjet > 54 \ \rm GeV/c$ and rapidity  in the region $0.1 < | \yjet | < 0.7$,
which are well described by NLO pQCD~\cite{jetrad} predictions \footnote{Previous measurements using the  $\kt$ algorithm at the
Tevatron~\cite{d0kt} observed a marginal agreement with NLO pQCD at low $\ptjet$ but 
this discrepancy is removed after non-perturbative corrections are included.}.
CDF has presented new measurements~\cite{cdfwww} of the inclusive jet production cross section  as a function of $\ptjet$
in five different jet rapidity regions up to $| \yjet | = 2.1$,
based on  $1.0 \ \rm fb^{-1}$ of CDF Run II data.

\begin{figure}
\centering
\includegraphics[width=8 cm]{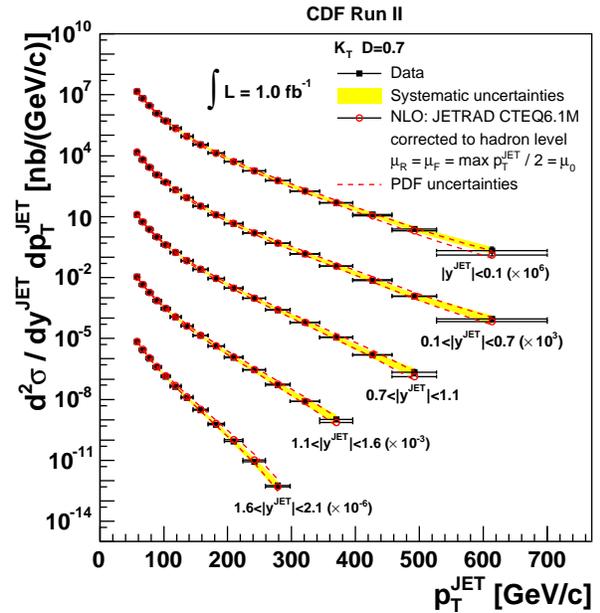}
\caption{
Measured inclusive differential jet cross sections, using the $\kt$ algorithm with $\rm D=0.7$, (black dots) as a function of $\ptjet$
for jets with $\ptjet > 54$~GeV/c
in different $|\yjet|$ regions compared to NLO pQCD predictions (open circles).
The shaded bands show the total systematic uncertainty on the measurements.
The dashed lines
indicate the PDF uncertainty on the theoretical predictions.
For presentation, each measurement is scaled by a given factor.} \label{xs}
\end{figure}
\begin{figure}
\centering
\includegraphics[width=8.5 cm]{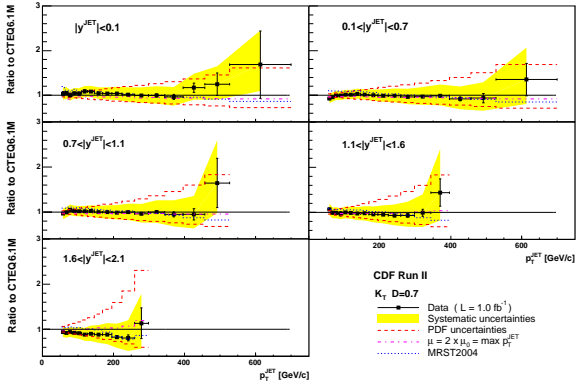}
\caption{Ratio Data/Theory as a function of $\ptjet$ in different $|\yjet|$ regions.
The error bars (shaded bands)
show the total statistical (systematic) uncertainty on the data.
A $5.8 \%$ uncertainty on the luminosity is
not included.
The dashed lines
indicate the PDF uncertainty on the theoretical predictions.
The dotted lines present the ratios of MRST2004  and CTEQ6.1M predictions.
The dotted-dashed lines show the ratios of predictions
with $2 \mu_0$ and $ \mu_0$.} \label{ratio}
\end{figure}

Figure~\ref{xs} shows the measured cross sections using the $\kt$ algorithm, 
\begin{equation}
\rm k_{\rm T,i} = p_{T,i}^2   \ \ \ \rm{;} \ \
\rm k_{\rm T, (i,j)} = \rm min(p_{T,i}^2,p_{T,j}^2) \cdot {\Delta R_{i,j}^2}/{D^2},
\end{equation}
\noindent
with $\rm D=0.7$,  as a function
of $\ptjet$ in five different $| \yjet |$ regions compared to NLO pQCD predictions where, for  presentation,
each measurement has been scaled by a given factor.
The measured cross sections decrease by more than seven
to eight orders of magnitude as $\ptjet$ increases.
Figure~\ref{ratio} shows the ratios data/theory as a function of $\ptjet$ in the five different $|\yjet|$ regions.
Good agreement is observed in the whole range in $\ptjet$ and  $\yjet$
between the measured cross sections  and the theoretical predictions.  In particular,
no significant deviation from the pQCD prediction is observed for central jets at high $\ptjet$.
In the most forward region, the uncertainty on the measured cross section  at high $\ptjet$,  compared to that on the
theoretical prediction, indicates that the data  will contribute to
a better understanding of the gluon~PDF.

\begin{figure}
\centering
\includegraphics[width=8.5 cm]{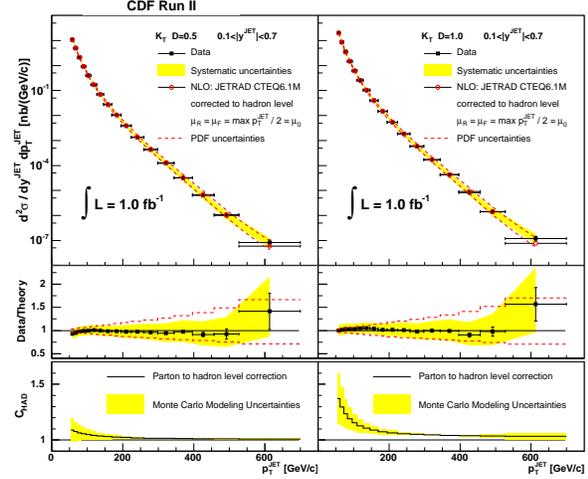}
\caption{Measured inclusive differential jet cross sections (black dots) as a function of $\ptjet$
for jets with $\ptjet > 54$~GeV/c
and $0.1 < |\yjet| < 0.7$ using $\rm D=0.5$ (left) and $\rm D=1.0$ (right), compared to NLO pQCD predictions.
The shaded bands show the total systematic uncertainty on the measurements.
A $5.8 \%$ uncertainty on the luminosity is
not included.
The dashed lines
indicate the PDF uncertainty on the theoretical predictions. (bottom) Magnitude of the parton-to-hadron corrections, 
$\rm C_{\rm HAD}(\ptjet)$,  used to correct
the NLO pQCD predictions for  $\rm D=0.5$ (left) and $\rm D=1.0$ (right). The shaded bands indicate the 
quoted Monte Carlo modeling uncertainty.} \label{xsd}
\end{figure}

In the region $0.1 < |\yjet| < 0.7$, the analysis is repeated using
different values for $\rm D$ in
 the $\kt$ algorithm: $\rm D=0.5$ and $\rm D=1.0$ (see Figure~\ref{xsd}).
In both cases, good agreement is observed between
 the measured cross sections and the NLO pQCD predictions in the whole range in $\ptjet$.
As $\rm D$ decreases, the measurement is less sensitive to contributions from multiple proton-antiproton
interactions, and the presence and proper modeling of the underlying event.  These measurements support the 
validity of the $\kt$ algorithm to search for jets in hadron collisions that will be further explored at the 
LHC energies.

\begin{figure}
\centering
\includegraphics[width=7 cm]{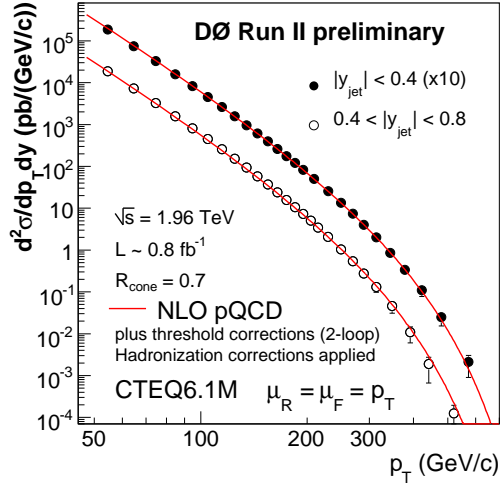}
\caption{Measured inclusive jet cross section as a function of $\ptjet$ in different $|\yjet|$ ranges (dots), 
compared to pQCD NLO predictions~\cite{d0nlo} (full lines). For presentation one of the measurements is scaled by a given 
factor.} \label{d0QCD1}
\end{figure}

\begin{figure}
\centering
\includegraphics[width=7 cm]{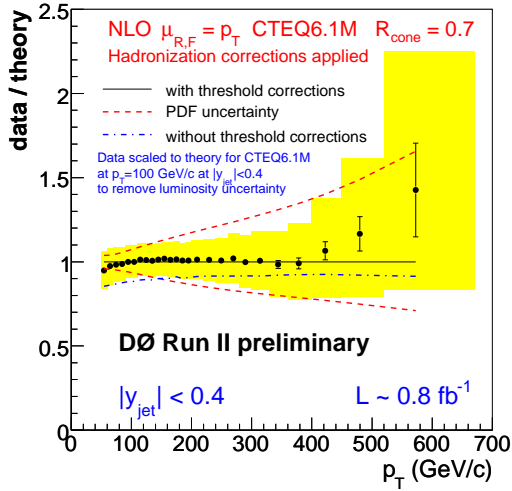}
\caption{Ratio data/theory as a function of $\ptjet$ for jets in the region $|\yjet|<0.4$. The band indicates
the uncertainty on the data and the dashed lines the uncertainty on the NLO prediction due to PDFs. The data is  
scaled to the theoretical prediction at $\ptjet = 100$~GeV/c.} \label{d0QCD2}
\end{figure}

Figure~\ref{d0QCD1} shows the measured inclusive jet cross section by D0~\cite{d0www}
based on the first 800~${\rm pb}^{-1}$ of Run II data. The new
midpoint  jet algorithm has been used with a cone size R=0.7. 
The measurements have been performed for central jets in two different $\yjet$ bins. 
Figure~\ref{d0QCD2} presents the ratio data vs NLO pQCD predictions~\cite{d0nlo} as a function of $\ptjet$ 
for jets in the region $|\yjet|< 0.4$. In order to eliminate the uncertainty on 
the integrated luminosity the ratio has been normalized to unity at $\ptjet = 100$~GeV/c.
The data is in good agreement with the  pQCD NLO predictions.

\section{UNDERLYING EVENT STUDIES}

As mentioned in the previous section, the hadronic final states at the
Tevatron are characterized by the presence of soft underlying
emissions, usually denoted as {\it{underlying
 event}}, in addition to highly energetic jets coming from the
hard interaction. The underlying event contains contributions from
initial- and final-state soft gluon radiation, secondary semi-hard partonic
interactions and interactions between the proton and anti-proton remnants that
cannot be described by perturbation theory. These processes  must be
approximately modeled using  Monte Carlo programs tuned to describe the data.

The jet energies measured in the detector contain an underlying event
contribution that has to be subtracted in order to compare the measurements
to pQCD predictions.
Hence, a proper understanding of this underlying event contribution
is crucial to reach the desired precision in
the measured jet cross sections.
In the analysis  presented here, the underlying
event  in dijet
events has been studied by looking at regions well separated from the leading
jets, where the underlying event contribution is expected to dominate the
observed hadronic activity. Jets have been reconstructed using  tracks with
$p_T^{\rm track} > 0.5$ GeV and $|\eta^{\rm track}|<1$ and a cone
algorithm with R=0.7.

\begin{figure}
\centering
\includegraphics[width=3.5 cm]{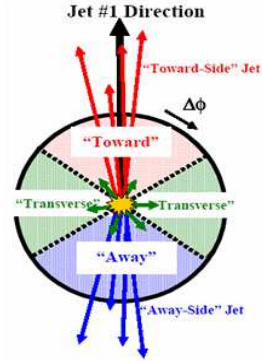}
\caption{Scheme of the different $\phi$ regions defined
around the leading jet.} \label{rf1}
\end{figure}

\begin{figure}
\centering
\includegraphics[width=8 cm]{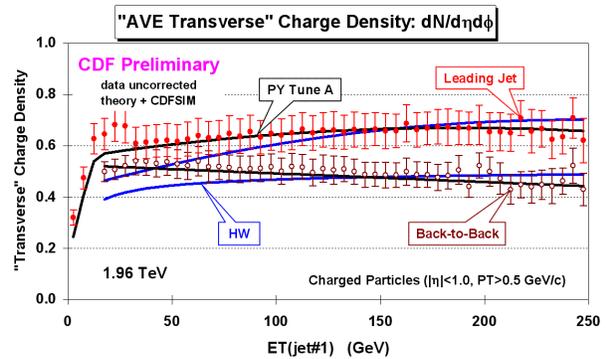}
\caption{Measured average track density
in the transverse region as a function of the $E_T^{\rm jet}$
of the leading jet. The measurements are compared to different Monte Carlo models.} \label{rf2}
\end{figure}

Figure~\ref{rf2} shows the average track density in the
transverse region as a function of $E_T^{\rm jet}$ of the leading jet for
the dijet inclusive sample and for events where the leading jets are forced
to be back-to-back in $\phi$, in order to further reduce extra hard-gluon radiation.
The observed plateau indicates that the underlying event activity
is, to a large extend, independent from the hard interaction.
The measurements have been compared to the
predictions from PYTHIA~\cite{pythia} and HERWIG~\cite{herwig} Monte Carlo programs
including  leading-order QCD matrix elements plus
initial and final parton showers. The PYTHIA samples have been created using a special tuned
set of parameters, denoted as PYTHIA-Tune A~\cite{tunea}, which includes
an enhanced contribution from initial-state soft gluon radiation and a tuned set 
of parameter to control secondary parton interactions. It was determined as a
result of similar studies of the underlying event performed using
CDF Run I data \cite{underlying}. PYTHIA-Tune A describes the hadronic activity
in transverse region  while HERWIG underestimates the radiation at low~$E_T^{\rm jet}$.

\section{JET SHAPES}

The internal structure of jets is dominated by  multi-gluon emissions from the 
primary final-state parton.
It is sensitive to the relative quark- and gluon-jet fraction  and
receives contributions from soft-gluon initial-state radiation and beam remnant-remnant interactions.
The study of jet shapes at the Tevatron provides a stringent test of QCD predictions  and tests
the validity of the models for parton cascades and soft-gluon emissions in hadron-hadron collisions.

\begin{figure}
\centering
\includegraphics[width=3.5 cm]{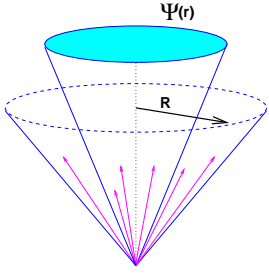}
\caption{Integrated jet shapes definition} \label{psi}
\end{figure}

The CDF experiment has presented results~\cite{shapes} on jet shapes for
central jets with transverse momentum in the region $37 < \ptjet < 380$~GeV,
where jets are searched for using the midpoint\footnote{A $75 \%$
merging fraction has been used instead of the default $50 \%$.} algorithm  and a cone size $R=0.7$.
The integrated jet shape, $\Psi(r)$, is defined as the average fraction of the
jet transverse momentum that lies inside a cone of radius $r$ concentric to the jet cone:
\begin{equation}
\Psi(r) = \frac{1}{\rm N_{jet}} \sum_{\rm jets} \frac{P_T(0,r) }{P_T(0,R)}, \ \ \ \ 0 \leq  r \leq R
\end{equation}
\noindent
where $\rm N_{\rm jet}$ denotes the number of jets. The measured jet shapes have been compared to the
predictions from PYTHIA-Tune A and HERWIG Monte Carlo programs.
\begin{figure}
\centering
\includegraphics[width=9.0 cm]{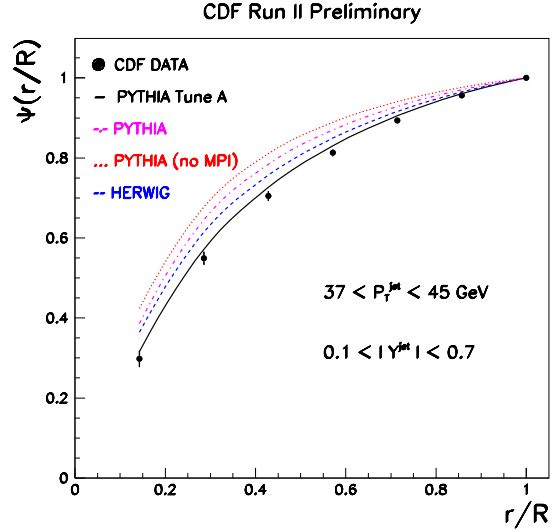}
\caption{The measured integrated jet shape, $\Psi(r/R)$, in inclusive jet production for jets
with $0.1 < |\yjet| < 0.7$ and $37 \ {\rm GeV/c} < \ptjet < 45 \ {\rm GeV/c}$.  
The predictions of PYTHIA-Tune~A 
(solid lines), PYTHIA (dashed-dotted lines), PYTHIA-(no MPI) (dotted lines) and HERWIG (dashed lines)
are shown for comparison.} \label{sh1}
\end{figure}
In addition, two different PYTHIA samples have been used with default parameters and with and without
the contribution from multiple parton interactions (MPI) between proton and antiproton remnants, the latter
denoted as PYTHIA-(no MPI), to illustrate the importance of a proper modeling of soft-gluon
radiation in describing the measured jet shapes. Figure~\ref{sh1}  presents the measured integrated 
jet shapes, $\Psi(r/R)$, for jets with $37 < \ptjet < 45$ GeV, compared to
HERWIG, PYTHIA-Tune A,  PYTHIA and PYTHIA-(no MPI) predictions. 

\begin{figure}
\centering
\includegraphics[width=9.0 cm]{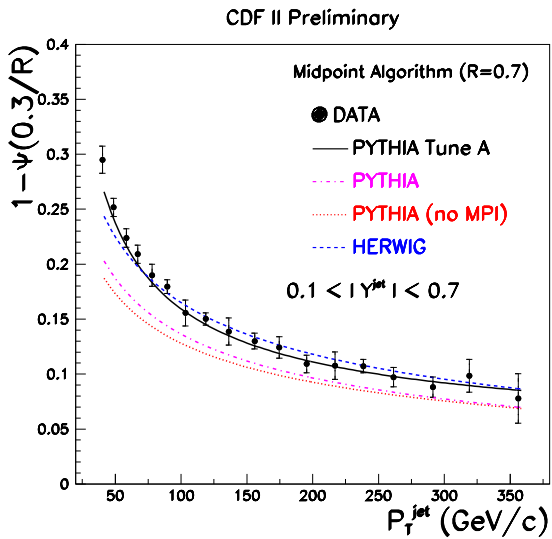}
\caption{The measured $1 - \Psi(0.3/R)$ as a function of $\ptjet$
for jets with $0.1 < |\yjet| < 0.7$ and $37 \ {\rm GeV/c} < \ptjet < 380 \ {\rm GeV/c}$.
Error bars indicate the statistical and systematic uncertainties added in quadrature.
The predictions of PYTHIA-Tune A (solid line)
, PYTHIA (dashed-dotted line),
PYTHIA-(no MPI) (dotted line) and HERWIG (dashed line)
are shown for comparison.} \label{sh2}
\end{figure}

Figure~\ref{sh2}
shows, for a fixed radius $r_0 = 0.3$, the average
fraction of the jet transverse momentum outside $r=r_0$, $1-\Psi(r_0/R)$, as a function of
$\ptjet$ where the points are located at the weighted mean in each $\ptjet$ range.
The measurements show that the fraction of jet transverse momentum at a given fixed $r_0/R$ increases
($1-\Psi(r_0/R)$ decreases) with $\ptjet$, indicating
that the jets become narrower as  $\ptjet$ increases. PYTHIA with default parameters
produces jets systematically narrower than the data in the whole region in $\ptjet$. 
The contribution from secondary parton interactions between remnants to the predicted jet shapes
(shown by the difference between  PYTHIA and PYTHIA-(no MPI) predictions) is relatively small
and decreases as $\ptjet$ increases. PYTHIA-Tune A predictions describe all of
the data well. HERWIG predictions describe
the measured jet shapes well for  $\ptjet > 55$ GeV  but
produces jets that are too narrow at lower $\ptjet$.

\section{DIJET AZIMUTHAL DECORRELATIONS}

The D0 experiment has employed the dijet sample to study azimuthal decorrelations, $\Delta \phi_{\rm dijet}$,  between the
two leading jets~\cite{dphi}. The normalized cross section,
\begin{equation}
\frac{1}{\sigma_{\rm dijet}} \frac{d \sigma}{d \Delta \phi_{\rm dijet}}, 
\end{equation}
is sensitive to the spectrum  of the gluon radiation in the event. The measurements has been performed in
different regions of the leading jet $\ptjet$ starting at $\ptjet > 75$~GeV,  where the second jet is
required to have at least $\ptjet > 40$~GeV.

\begin{figure}
\centering
\includegraphics[width=6 cm]{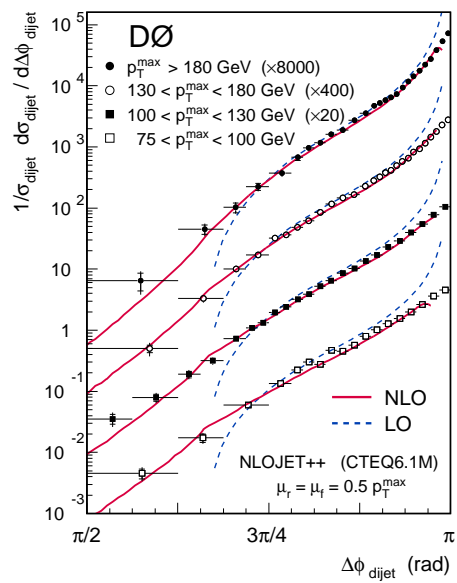}
\caption{ Measured azimuthal decorrelations in dijet production for central jets 
compared to pQCD predictions in
different regions of $\ptjet$ of the leading jet.} \label{dph1}
\end{figure}

Figure~\ref{dph1} shows the measured cross section compared to LO and NLO predictions~\cite{dphinlo}.
The LO predictions, with at most three partons in the final state, is limited to $\Delta \phi_{\rm dijet} > 2 \pi/3$,
for which the three partons define a {\it{Mercedes-star}} topology. It presents a prominent peak
at $\Delta \phi_{\rm dijet} = \phi$ corresponding to the soft limit for which the third parton is collinear
to the direction of the two leading partons. The NLO predictions, with four partons in the final state, describes
the measured  $\Delta \phi_{\rm dijet}$ distribution except at very high and very low values of  $\Delta \phi_{\rm dijet}$
where additional soft contributions, corresponding to a resummed calculation, are necessary. A reasonable approximation
to such calculations is provided by parton shower Monte Carlo programs.

\begin{figure}
\centering
\includegraphics[width=6 cm]{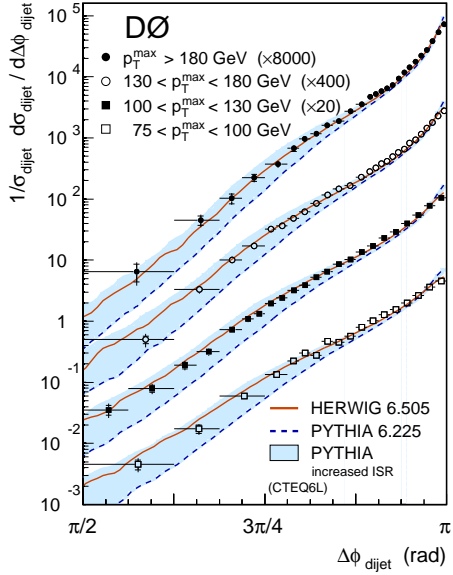}
\caption{ Measured azimuthal decorrelations in dijet production for central jets compared to PYTHIA and HERWIG
 predictions in different regions of leading $\ptjet$. The band covers
PYTHIA predictions with different amount of initial-state soft-gluon radiation.} \label{dph2}
\end{figure}

Figure~\ref{dph2} present the measured cross section compared to PYTHIA and HERWIG predictions 
in different regions of $\ptjet$. PYTHIA with default parameters underestimates the gluon radiation at large angles.
Different tunes of PYTHIA predictions are possible, which include an enhanced contribution from initial-state 
soft gluon radiation, to properly describe the azimuthal distribution. HERWIG also describes the data although 
tends to produce less radiation than PYTHIA close to the direction of the leading jets. 
This measurements clearly show that angular correlations 
between jets can be employed to tune Monte Carlo predictions of 
soft gluon radiation in the final state. 

\section{DIRECT PHOTON PRODUCTION}

\begin{figure}
\centering
\includegraphics[width=8 cm]{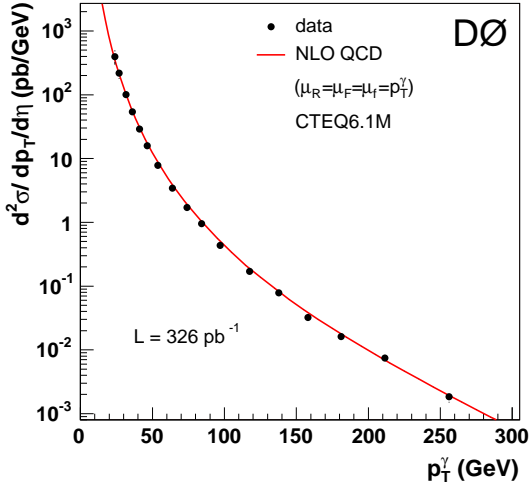}
\caption{Measured inclusive prompt photon cross section as a function of $\ptg$ (dots) compared 
to pQCD NLO prediction (solid line).} \label{gamma1}
\end{figure}

The measurement of the inclusive  photon production as a function of photon transverse
momentum, $\ptg$, at the Tevatron constitutes a precise test of pQCD predictions since 
the energy scale of the electromagnetic calorimeters is well understood by the experiments. 
The measured cross section  is partially dominated by contributions from quark-gluon scattering and
therefore provides a powerful constrain of the gluon PDF at high-x.  However, it is a 
rather difficult measurement where a good understanding of QCD backgrounds from 
$\pi^0$ and $\eta$ decays into photons is necessary. The D0 collaboration has
presented results~\cite{photon} on inclusive photon production, based on 326 $\rm pb^{-1}$  of Run II data, 
in the region $|\eta^{\gamma}| < 0.9$ and $\ptg < 300$~GeV/c (see Figure~\ref{gamma1}). The measurements
are compared to NLO pQCD predictions~\cite{gnlo}. 

\begin{figure}
\centering
\includegraphics[width=8 cm]{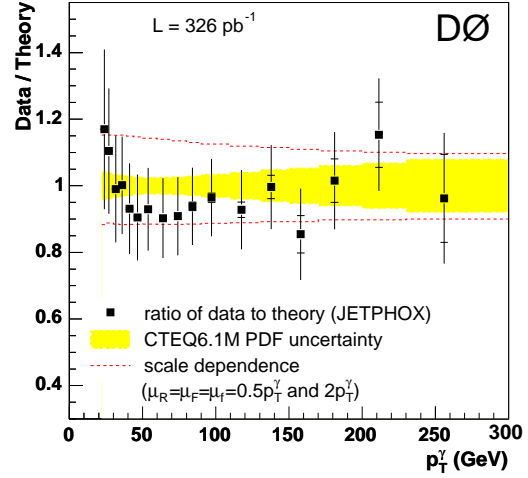}
\caption{Ratio data/theory as a function of $\ptg$. The error bars show the uncertainty 
on the measurements. The band denotes the PDF uncertainty on the theoretical prediction and 
the dashed lines indicate the uncertainty from the variation of the renormalization/factorization 
scales in the pQCD NLO calculation.} \label{gamma2}
\end{figure}

Figure~\ref{gamma2} shows the ratio data/NLO as a
function of $\ptg$. The measured cross section is well described by the theoretical prediction, where
the latter presents uncertainties at the level of about $10 \%$.  Future measurements based on 
few $\rm fb^{-1}$ of data will provide valuable information about the proton structure as well as 
imposes strong constrains to the presence of new physics with very energetic photons in the final state.

\section{BOSON PLUS JETS PRODUCTION}

\begin{figure}
\centering
\includegraphics[width=8 cm]{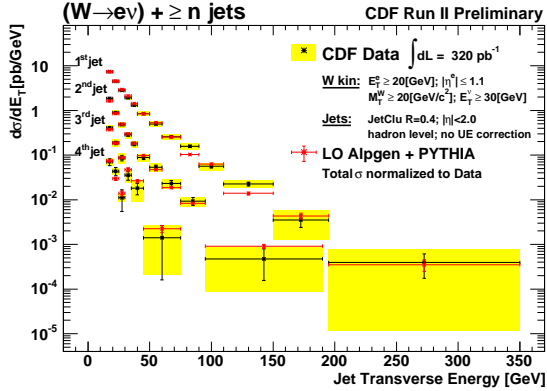}
\caption{Measured inclusive cross section as a function of $\etjet$ for the n${}^{th}$ jet in 
inclusive $\rm W+n_{\rm jet}$ production. The measurements are compared to different leading-order Monte Carlo 
predictions normalized to the data.} \label{cdfW1}
\end{figure}

\begin{figure}
\centering
\includegraphics[width=8.0 cm]{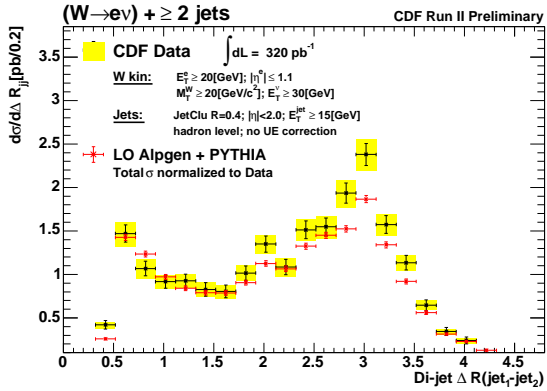}
\caption{Measured cross section as a function of $\Delta R_{\rm jet1-jet2}$ in inclusive $\rm W+ \geq 2$ jets production 
compared to leading-order Monte Carlo predictions. The Monte Carlo distribution is normalized to the 
data. } \label{cdfW2}
\end{figure}

The study of the production of electroweak bosons in association with jets of hadrons in the final
state   constitutes a fundamental item in the high-$\pt$ physics program at the Tevatron. These events 
are main backgrounds to many interesting physics processes like, for example, top production, and the search
for the SM Higgs and supersymmetry.  Therefore, during the last few years a significant effort is being made 
to develop and validate the necessary Monte Carlo tools to describe these complicated multijet 
final states. As a result, a number of leading-order  Monte Carlo programs are available  that
describe boson+jet production processes up to large parton multiplicities~\cite{vecbos,alpgen,madgraph}, 
and NLO pQCD parton-level predictions are also available for a limited number of processes 
(up to  boson+2jets production)~\cite{mcfm}. The interface
between parton level calculations and  parton showers, necessary to describe the complexity of the observed 
 hadronic final states,  requires the introduction of different prescriptions  
to resolve resulting double counts across processes with different parton multiplicities.  The 
theoretical prescriptions employed require validation using data.  For this purpose, both CDF and D0 collaborations have performed 
a careful set of measurements on Boson+jets production. Figure~\ref{cdfW1} shows
the measured $\ptjet$ distribution for the  ${\rm n}^{th}$ jet  in inclusive $\rm W+n_{\rm jet}$ production 
by CDF, based on 320 $\rm pb^{-1}$ of 
Run II data.  This observable is particularly sensitive
to the details on the implementation of the parton shower interface in the Monte Carlo models. The measurements are compared to 
leading-order  Monte Carlo~\cite{alpgen} predictions interfaced with PYTHIA parton shower, and normalized 
to the data. Similarly, Figure~\ref{cdfW2} shows the measured distance, ($\eta - \phi$) space, between the two 
leading jets in inclusive $\rm W +  2 \rm jets$
production. The Monte Carlo model provides a reasonable description of the shape of the measured distributions.  

\begin{figure}
\centering
\includegraphics[width=8 cm]{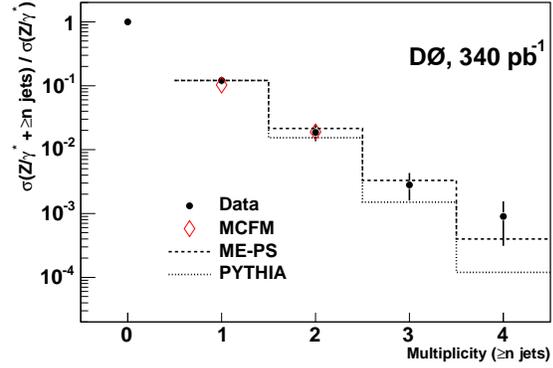}
\caption{Measured inclusive cross section as a function of jet multiplicity for $Z+N_{\rm jet}$ production.
The measurements are normalized to the total Drell-Yan production cross section. The data is compared
to different leading-order hadron level Monte Carlo and NLO pQCD parton level predictions. ME+PS predictions 
are normalized to the measured inclusive $Z+1{\rm jet}$ production yield.} \label{d0Z1}
\end{figure}

\begin{figure}
\centering
\includegraphics[width=8 cm]{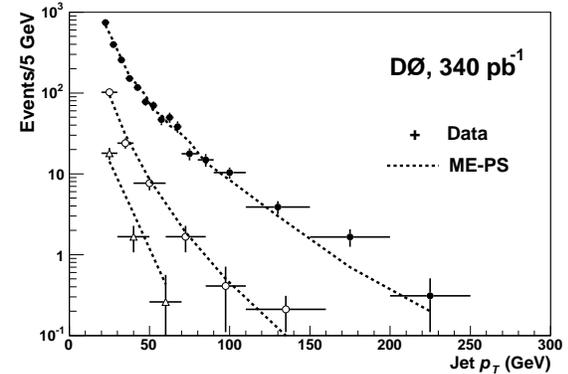}
\caption{Measured $\etjet$ distribution for the n${}^{th}$ jet in 
$\rm W+N_{\rm jet}$ production. The measurements are compared to different leading-order Monte Carlo 
predictions.} \label{d0Z2}
\end{figure}

The D0 collaboration has performed a detailed measurement on inclusive 
$\rm Z +n_{\rm jet}$ production~\cite{d0Zs}. Figure~\ref{d0Z1} presents
the measured cross section, normalized to the total Drell-Yan cross section, as a function of the inclusive jet multiplicity.  
The measurements are compared to PYTHIA, Matrix Elements+Parton Shower (ME+PS)~\cite{madgraph}, and parton-level 
NLO calculations~\cite{mcfm}. 
As expected, PYTHIA, that only includes matrix elements for two-to-two processes, only 
provides a reasonable description for one and two inclusive jet 
production\footnote{The hardness of the first gluon radiation in a parton showers approximately follows that of matrix elements.},  
while underestimates the production of events with large jet multiplicities.  The prediction from ME+PS describes 
the observed normalized yields as well as the shape of the $\rm n^{\rm th}$ jet $\ptjet$ distributions 
(see Figure~\ref{d0Z2}). Finally,  pQCD NLO predictions for one and two inclusive jet production provide a 
reasonable description of the data. 

\begin{figure}
\centering
\includegraphics[width=7 cm]{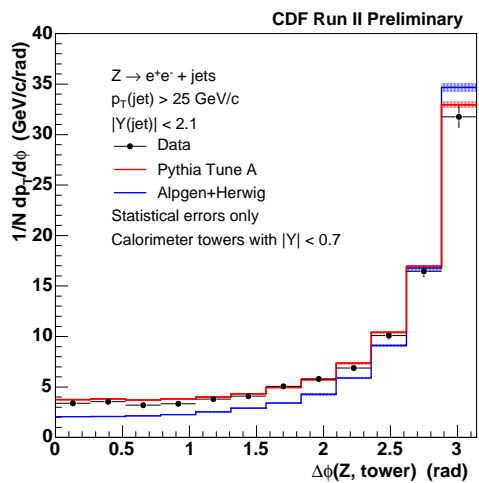}
\caption{Measured mometum flow (transverse plane) and a function of $\Delta \phi$ in inclusive Z+jet production, where
the Z direction defined $\phi = 0$. The measurements are compared with PYTHIA and ALPGEN+HERWIG.} \label{cdfZ1}
\end{figure}

\begin{figure}
\centering
\includegraphics[width=7 cm]{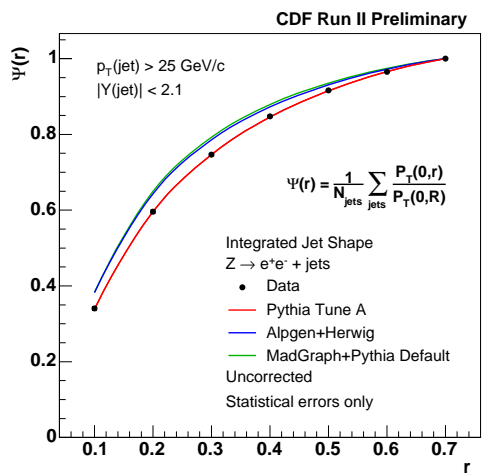}
\caption{Measured integrated jet shape in inclusive Z+jet production for jets in the region $\ptjet > 25$~GeV/c and 
$|\yjet|<2.1$, compared to different leading-order Monte Carlo predictions.} \label{cdfZ2}
\end{figure}

The CDF collaboration has presented first measurements of jet shapes and energy flows in inclusive Z+jet production, which 
are necessary to validate the modeling for underlying event and soft gluon radiation implemented in the Monte Carlo generators. 
Figure~\ref{cdfZ1} shows the measured momentum flow (projected to the transverse plane) as a function of the azimuthal distance
with respect to the Z direction ($\phi = 0$), where only the central region of the calorimeter ($|\eta|<0.7$) is considered.
In the region $|\Delta \phi| = \phi$ the distribution shows the presence of the leading jet. The region $|\Delta \phi| \sim 1$
is particularly sensitive to a proper description of the underlying event. PYTHIA-Tune A provides a good description of the 
measured transverse momentum flow while ALPGEN+HERWIG would require additional underlying event activity.    
Similar conclusions can be drawn from Figure~\ref{cdfZ2}, where the measured jet shape in inclusive Z+jet production is presented.
PYTHIA Tune A provides the best description of the data, and the difference between models can be attributed to differences
on the underlying event implementation.   

\section{FINAL NOTES}
The Tevatron is delivering luminosity according to expectations and each experiment plans to collect 
more than $\rm 4 \  fb^{-1}$ of data. During the next few year the jet physics program at the Tevatron 
will translate into a number of precise measurements that will test the SM and provide very valuable 
information on how to model QCD processes at the LHC. 

\bigskip 
\begin{acknowledgments}
I would like to thank the organizers for their kind invitation to the conference.   
\end{acknowledgments}

\bigskip 


\begin{thebibliography}{9}   


\bibitem{d0runI} B. Abbott {\it et al.}   (D\O  \ Collaboration), Phys. Rev. Let
t. {\bf 82}, 2451 (1999).\\
\bibitem{runIjet} T. Affolder {\it et al.} (CDF Collaboration), Phys. Rev. D {\b
f 64}, 032001~(2001).
[Erratum-ibid. D {\bf 65}, 039903 (2002)].

\bibitem{ktprl} A. Abulencia~{\it et al.} (CDF Collaboration), Phys. Rev. Lett. {\bf 96}, 122001 (2006).
\bibitem{runIIjet} A. Abulencia~{\it et al.}, hep-ex/0512020.



\bibitem{ktalgo}  S. Catani {\it et al.}, Nucl. Phys. B {\bf 406}, 187 (1993).
\bibitem{soper} S.D. Ellis and D.E. Soper,  Phys. Rev. D {\bf 48}, 3160 (1993).
\bibitem{midpoint}   G. C. Blazey, et al., hep-ex/0005012 (2000). \\
                     S.D. Ellis, et al.,  hep-ph/0111434 (2001).

\bibitem{jetrad}  W.T. Giele, E.W.N. Glover and David A. Kosower, Nucl. Phys. B {\bf 403}, 633 (1993).

\bibitem{d0kt} V.M. Abazov {\it et al.}   (DO Collaboration), Phys. Lett. B {\bf 525}, 211 (2002).
\bibitem{cdfwww} see http://www-cdf.fnal.gov

\bibitem{d0www} see http://www-d0.fnal.gov
\bibitem{d0nlo} N. Kidonakis and J.F. Owens, Phys. Rev. D 63, 054019 (2001), hep-ph/0007268 

\bibitem{pythia} T. Sj\"ostrand {\it et al.}, Comp. Phys. Comm. {\bf 135}, 238 (2001).
\bibitem{herwig} G. Corcella {\it et al.}, JHEP {\bf 0101}, 010 (2001).
\bibitem{tunea} {\sc pythia-tune~a}  Monte Carlo samples are generated using the following tuned parameters in {\sc pyth
ia}:
   PARP(67) = 4.0, MSTP(82) = 4, PARP(82) = 2.0, PARP(84) = 0.4, PARP(85) = 0.9, PARP(86) = 0.95, PARP(89) = 1800.0, PAR
P(90) =
 0.25.
\bibitem{underlying} T.~Affolder {\it et al.} (CDF Collaboration), Phys. Rev. D {\bf 65}, 092002 (2002).
\bibitem{shapes}  D. Acosta {\it et al.}    (CDF Collaboration), Phys. Rev. D {\bf 71}, 112002 (2005).
\bibitem{dphi} V.M. Abazov {\it et al.}     (D0 Collaboration), Phys. Rev. Lett. {\bf 94}, 221801 (2005).
\bibitem{dphinlo} Z. Nagy, Phys. Rev. Lett. {\bf 88}, 122003 (2002). 
\bibitem{photon} V.M. Abazov {\it et al.}     (D0 Collaboration), Phys. Lett.  B {\bf 639} 151 (2006).
\bibitem{gnlo} T. Binoth et al., Eur. Phys. J. C{\bf 16}, 311 (2000).\\
               S. Catani et al., JHEP {\bf 05}, 028 (2002).
\bibitem{vecbos} F.A. Berends {\it et al.},   Nucl Phys B {\bf 357} 32 (1991).
\bibitem{alpgen} M.L. Mangano {\it et al.}, JHEP {\bf 0307} 001 (2003).
\bibitem{madgraph} F. Maltoni and T. Stelzer, JHEP {\bf 0302} 027 (2003). 
\bibitem{mcfm} J. Campbell, R.K. Ellis and D. Rainwater, Phys. Rev. D {\bf68} 094021 (2003).
\bibitem{d0Zs} V.M. Abazov {\it et al.}     (D0 Collaboration), FERMILAB-PUB-06-283-E.

\end{thebibliography}
\end{document}